\documentclass[twocolumn,showpacs,preprintnumbers,nofootinbib,amsmath,amssymb,superscriptaddress]{revtex4}
\usepackage{epsfig} %Inlcude figure files
\usepackage{graphicx}% Include figure files
\usepackage{dcolumn}% Align table columns on decimal point
\usepackage{bm}% bold math

\newcommand{\bee}{\begin{equation}}
\newcommand{\ee}{\end{equation}}
\newcommand{\beea}{\begin{eqnarray}}
\newcommand{\eea}{\end{eqnarray}}

\begin{document}

\preprint{SNUTP-04-008}
\preprint{UW/PT 04-11}
\preprint{LA-UR-04-4788}

\title{
Testing improved staggered fermions with $m_s$ and $B_K$ 
}
\author{Weonjong Lee}
\email{wlee@phya.snu.ac.kr}
\affiliation{
  School of Physics,
  Seoul National University,
  Seoul, 151-747,
  South Korea
}
\author{Tanmoy Bhattacharya}
\email{tanmoy@lanl.gov}
\affiliation{
MS--B285, T-8,
Los Alamos National Lab,
Los Alamos, New Mexico 87545, USA
}
\author{George T. Fleming}
\email{flemingg@jlab.org}
\affiliation{
Jefferson Lab, 
12000 Jefferson Avenue, 
Newport News,
VA 23606, USA
}
\author{Rajan Gupta}
\email{rajan@lanl.gov}
\affiliation{
MS--B285, T-8,
Los Alamos National Lab,
Los Alamos, New Mexico 87545, USA
}
\author{Gregory Kilcup}
\email{kilcup@physics.ohio-state.edu}
\affiliation{
  Department of Physics,
  Ohio State University, 
  Columbus, OH 43210, USA
}
\author{Stephen R. Sharpe}
\email{sharpe@phys.washington.edu}
\affiliation{
  Physics Department,
  University of Washington,
  Seattle, WA 98195-1560, USA
}
\date{\today}
\begin{abstract}
We study the improvement of staggered fermions using hypercubically
smeared (HYP) links. We calculate the strange quark mass and the kaon
B-parameter, $B_K$, in quenched QCD on a $16^3 \times 64$ lattice at
$\beta=6.0$. We find $m_s(\overline{\rm MS},2\;{\rm
GeV})=101.2\pm1.3\pm4\;$MeV and $B_K(\overline{\rm MS},2\;{\rm GeV}) =
0.578 \pm 0.018\pm 0.042$, where the first error is from statistics
and fitting, and the second from using one-loop matching factors.
The scale  ($1/a=1.95$GeV) is set by $M_\rho$,
and $m_s$ is determined using the kaon mass.  
Comparing to quenched results obtained using unimproved staggered
fermions and other discretizations, we argue that the size of
discretization errors in $B_K$
is substantially reduced by improvement.
\end{abstract}
\maketitle

\section{Introduction}
\label{sec:intr}

%--------------
% introduction
%--------------
%
% B_K
%
%
One of the major sources of uncertainty in using precision
experimental data to constrain the standard model is the lack of
knowledge of the matrix elements of the 
effective weak Hamiltonian between hadronic states.
The kaon bag parameter $B_K$, which parameterizes 
the matrix element of the
$\Delta S = 2$ operator responsible for kaon-antikaon mixing, is one
such key input for the determination of the CKM mixing matrix.
It is defined as the dimensionless ratio 
\begin{equation}
B_K = \frac{\langle \bar{K}^0 | \bar{s}\gamma_\mu(1-\gamma_5)d 
  \ \bar{s}\gamma_\mu(1-\gamma_5)d | K^0 \rangle}
{ \frac{8}{3} \langle \bar{K}^0 | \bar{s}\gamma_\mu \gamma_5 d | 0 \rangle
 \langle 0 | \bar{s}\gamma_\mu \gamma_5 d | K^0 \rangle }
\label{eq:bkdef}
\end{equation}
Different approaches, including chiral perturbation theory, the large
$N_c$ expansion, QCD sum rules and lattice QCD, have been used to estimate
$B_K$.
The advantage of the lattice approach is that it is a first
principle, non-perturbative determination. On the other hand it
introduces statistical and systematic errors like those due to
discretization and the matching of lattice and continuum operators.
%
%The dominant systematic errors in $B_K$ come from quenching or
%partial quenching, mixing with operators of wrong chirality,
%discretization errors, and the matching of lattice
%and continuum operators. 
%
To gain control over these uncertainties, different fermion
discretizations---Wilson, staggered, domain-wall (DW) and
overlap---have been used in simulations.\footnote{% 
See
Ref.~\cite{ref:becirevic:0} for a recent review.}

In this note we explore the extent to which improved staggered
fermions can be used to reduce two of the most important systematic
errors, $i.e.$, those due to discretization and the matching of
lattice and continuum operators. This test is carried out in the
quenched approximation to get an estimate of the size of these errors
by comparing with existing data.  Our ultimate aim, however, is to
find a method which can be used effectively on dynamical lattices
likely to be produced in the near future.

%%%%%%%%%%%%%%%%%%%%%
% Why using improved staggered fermions?
%%%%%%%%%%%%%%%%%%%%%

Staggered fermions are an attractive choice for the calculation of
weak matrix elements because they are computationally
efficient---indeed, simulations with three dynamical flavors are
already possible with relatively light quark
masses~\cite{ref:davies:0}--- and yet retain sufficient chiral
symmetry to protect operators of physical interest from mixing with
others of wrong chirality.
% and prevent additive ultra-violet 
% corrections to mass renormalization. 
%
Their disadvantage is that they retain four ``tastes'' of doublers for
each lattice field.  In the continuum limit, these four tastes become
degenerate, and one can remove the additional degrees of freedom by
hand. For the valence quarks this procedure is explained for the
calculation of $B_K$ in Ref.~\cite{ref:smearedops}, while for the sea
quarks one must take the fourth-root of the quark determinant.  At
non-zero lattice spacing, however, quark-gluon interactions violate
the taste symmetry.  This has three important consequences for
calculations of $B_K$.

The first concerns taste symmetry violation and 
the need to take the fourth-root of the quark determinant.
For non-zero lattice spacing, there is no proof that the underlying
lattice action is local and lies in the same universality class as
QCD.  Even though we do not face this problem in quenched simulations,
it is relevant when extending our calculations to dynamical
simulations. Our justification for proceeding is empirical--- 
accurate unquenched simulations using the fourth-root of the
determinant find agreement between lattice and experimental
results~\cite{ref:davies:0}.

Second, large ${\cal O}(a^2)$ discretization errors have been observed
in the calculation of masses and matrix elements. Overcoming these
requires the use of very small lattice spacings to make reliable
continuum extrapolations. Lastly, many one-loop perturbative estimates
of matching factors differ significantly from their tree-level value
of unity, raising doubts about their accuracy~\cite{ref:Golterman}.

The purpose of this paper is to show, using $m_s$ and $B_K$ as probes,
that the latter two problems can be greatly alleviated by improving
staggered fermions using ``fat'' links~\cite{ref:lagae_lepage}.  Based
on the analysis of Ref.~\cite{ref:wlee:6}, we choose a particular type
of fattening, hypercubic (HYP) smeared links~\cite{ref:anna:1},
although we expect that other choices will work comparably well.
Earlier calculations show that taste-symmetry violations in the
spectrum are substantially reduced~\cite{ref:OT,ref:anna:1}, and
one-loop corrections to matching factors for four-fermion operators
which were as large as 100\% are now reduced to $\sim 10\%$
\cite{ref:wlee:4}. The largest improvement is in renormalization
constants of left-right (penguin) four-fermion
operators~\cite{ref:wlee:4}, and this can be traced back to the
improvement in $Z_m = 1/Z_S = 1/Z_P \approx 1$ with HYP smearing. This has a
major impact on the extraction of $m_s$ as we show in
section~\ref{sec:quark mass}. In the case of $B_K$ the major
impact of improvement is to reduce discretization errors.
This is because the one-loop corrections to matching factors 
in this case turn out to be small ($\sim 10\%$) before 
(as well as after) improvement.

To test the efficacy of improvement for $m_s$ and $B_K$, we compare
our results to the JLQCD analyses with unimproved staggered fermions
that include detailed studies of both discretization and perturbative
errors~\cite{ref:aoki:0,ref:aoki:1}.  

This paper is organized as follows. 
In Sec.~\ref{sec:mrho}, we analyze the $\rho$ meson spectrum calculated
on the HYP smeared lattices and obtain the lattice scale $1/a$.
In Sec.~\ref{sec:quark mass}, we present the extraction of strange
quark mass from the pion spectrum and compare the result with that
obtained using unimproved staggered fermions.
In Sec.~\ref{sec:B_K}, we present results for $B_K$ calculated using
the HYP improved staggered fermions and compare them with
those of unimproved staggered fermions and with some recent data 
obtained using domain wall and overlap fermion formulations.
We close with some conclusions in Sec.~\ref{sec:conclude}.

%This result has been confirmed and improved by
%a more recent calculation using DW fermions~\cite{ref:cp-pacs:1}.
%Using only two moderately small
%lattice spacings ($1/a\approx 1.8$ and $2.8\;$GeV) they 
%$B_K({\rm NDR},2\;{\rm GeV})=0.575 \pm 0.020$
%The largest two components in this error are from the use of one-loop
%matching and the continuum extrapolation.

%------------------------
% simulation parameters.
%------------------------
%
%
%
%% \begin{table}[t!]
%% \caption{Parameters for numerical study
%% \label{tab:param}}
%% \begin{ruledtabular}
%% \begin{tabular}{c | r}
%%   parameter  &  value \\ 
%% \hline
%%   gauge action    &  Wilson Plaquette \\
%%   fermion action  &  HYP staggered fermion \\
%%   lattice volume  &  $16^3 \times 64$ \\
%%   $\beta$         &  6.0 \\
%%   number of configurations & 218 \\
%% \end{tabular}
%% \end{ruledtabular}
%% \end{table}
%
%
%
%
\begin{table}[t!]
\caption{Quark masses used in simulation and their relation
to the strange quark mass.}
\label{tab:quarkmass}
\begin{ruledtabular}
\begin{tabular}{c | l l}
Name &  $a m_q$  & $m_q/m_s$ \\
\hline
  $m_1$  &  0.01 &  $0.192$ \\
  $m_2$  &  0.02 &  $0.385$ \\
  $m_3$  &  0.03 &  $0.577$ \\
  $m_4$  &  0.04 &  $0.769$ 
%\hline 
%  $m_s$  &  0.05202(67) &  % $101.6 \pm 1.3$ MeV 
\end{tabular}
\end{ruledtabular}
\end{table}

\section{$\rho$ meson spectrum}
\label{sec:mrho}

\begin{figure}[t!]
\includegraphics[width=20pc]{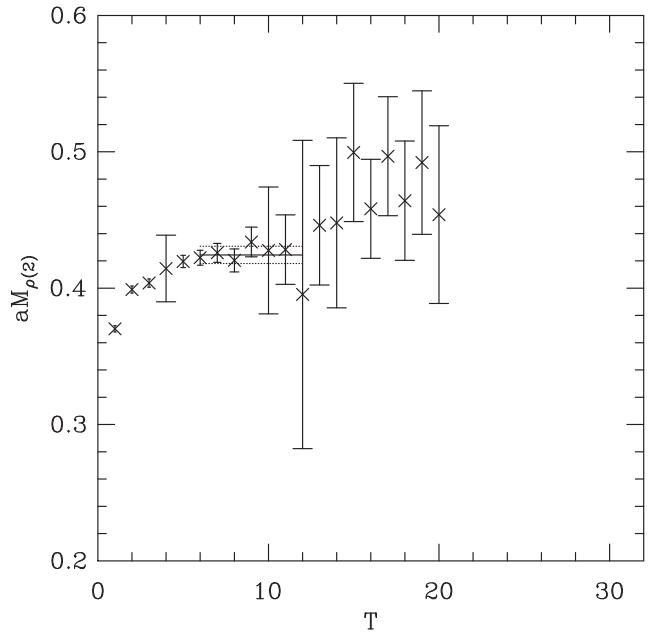}
%%%\vspace*{-5mm}
\caption{Effective mass plot of $aM_\rho$ at quark mass 0.01.
}
\label{fig:mrho_m1}
%%%\vspace*{-3mm}
\end{figure}

\begin{figure}[t!]
\includegraphics[width=20pc]{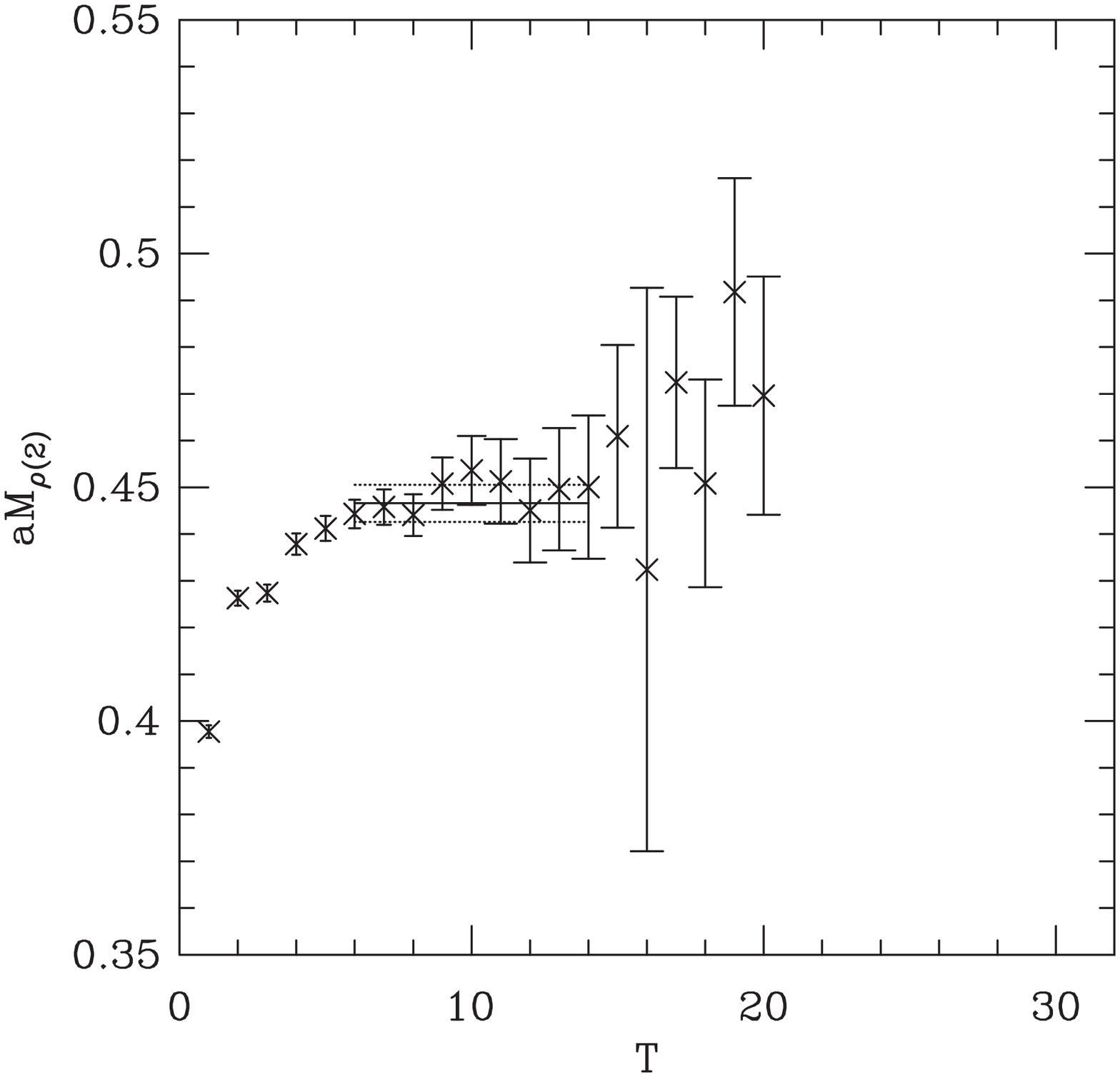}
%%%\vspace*{-5mm}
\caption{Effective mass plot of $aM_\rho$ at quark mass 0.02.
}
\label{fig:mrho_m2}
%%%\vspace*{-3mm}
\end{figure}

\begin{figure}[t!]
\includegraphics[width=20pc]{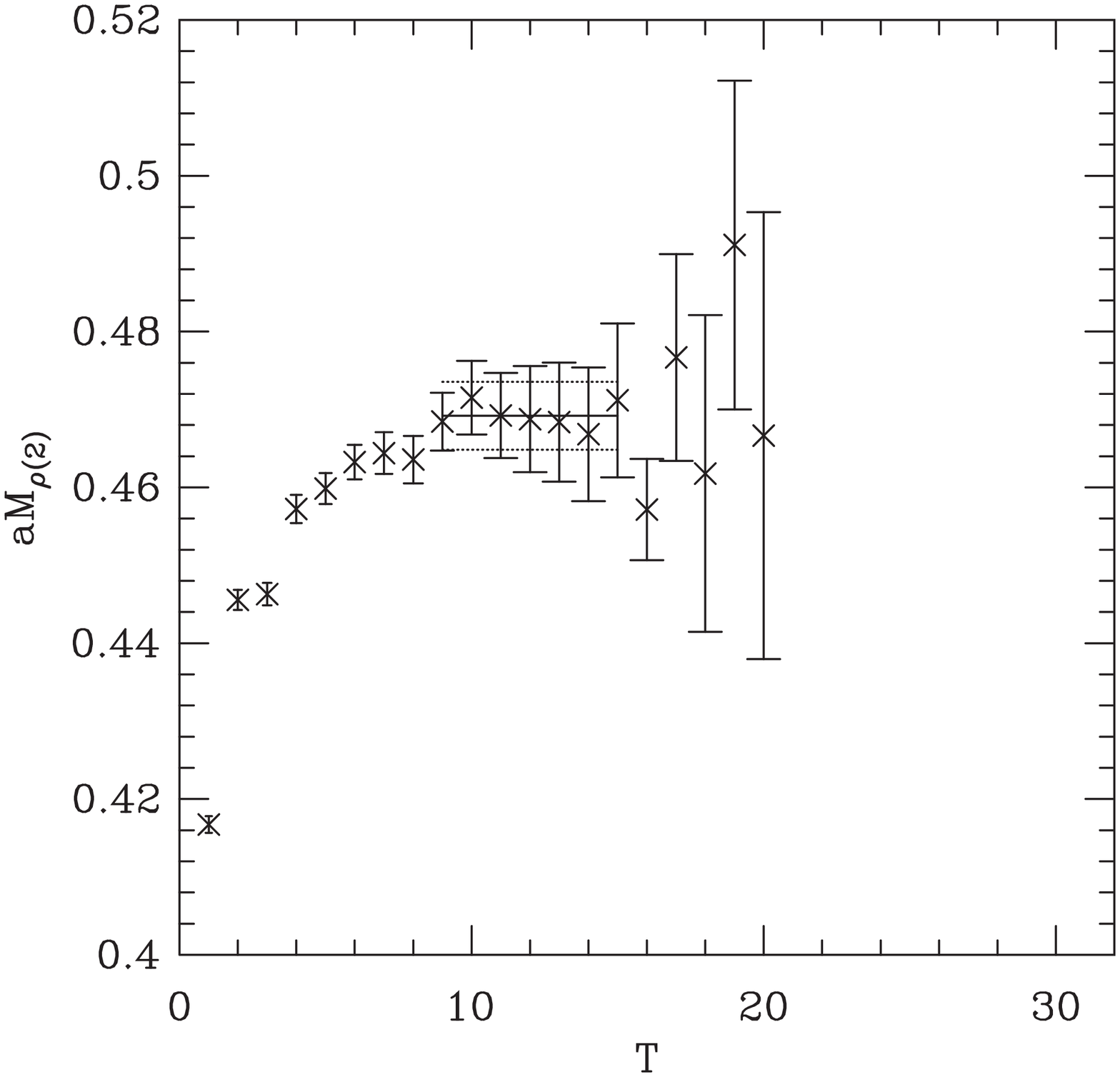}
%%%\vspace*{-5mm}
\caption{Effective mass plot of $aM_\rho$ at quark mass 0.03.
}
\label{fig:mrho_m3}
%%%\vspace*{-3mm}
\end{figure}

\begin{figure}[t!]
\includegraphics[width=20pc]{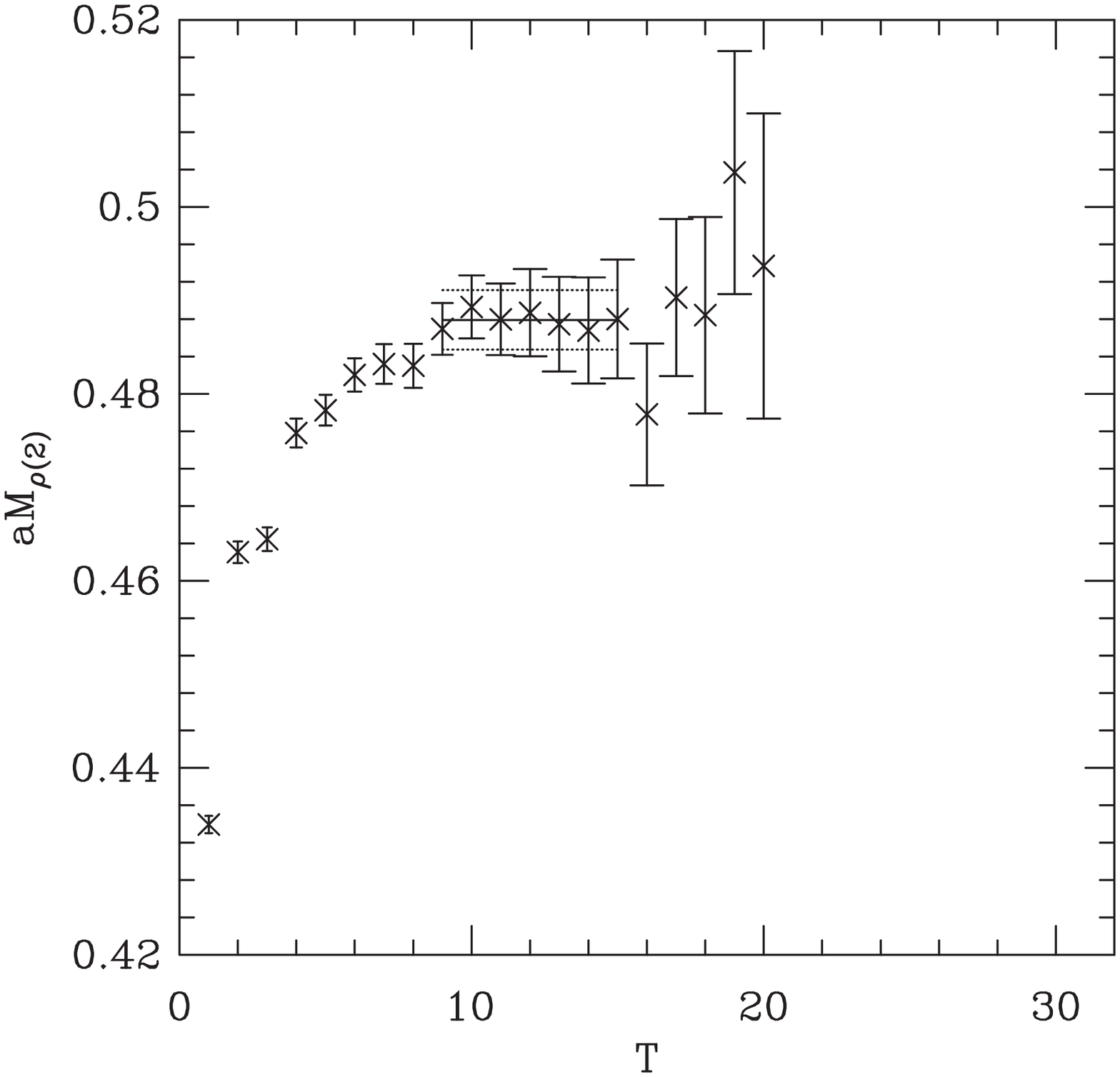}
%%%\vspace*{-5mm}
\caption{Effective mass plot of $aM_\rho$ at quark mass 0.04.
}
\label{fig:mrho_m4}
%%%\vspace*{-3mm}
\end{figure}

The statistical sample consists of an ensemble of 218 gauge
configurations of size $16^3 \times 64$ generated using the Wilson
plaquette action at $\beta=6.0$.  The lattices were first HYP smeared
using the tree-level improved parameters of Ref.~\cite{ref:wlee:6}.
On these HYP smeared lattices quark propagators are calculated using
$2Z$ wall sources on time-slices $0$, $16$, $32$ and $48$ for the bare
quark masses listed in Table \ref{tab:quarkmass}.
\footnote{The details on the $2Z$ wall source are given in
Ref.~\cite{ref:sui:1}.}
Meson correlators are calculated at all time slices for each set of
the $2Z$ wall sources.
Throughout this work we only consider mesons composed of degenerate
quarks.
The lattice scale is set using the $\rho$ meson mass. 
We calculated correlators for two types of $\rho$ mesons: $\rho(1)$,
with spin-taste 
$(\gamma_i \otimes \xi_i)$; and $\rho(2)$, with spin-taste 
$(\gamma_i\gamma_4 \otimes \xi_i\xi_4)$.
The correlators are fit to the standard form~\cite{ref:golterman:1}:
\begin{eqnarray}
C(t) &=& Z_1 \left\{ \exp\left[- m_1 t\right]
 + \exp\left[- m_1 (L-t)\right] \right\} 
\nonumber \\
     &+& Z_2 (-1)^t \left\{
 \exp\left[- m_2 t\right] + \exp\left[- m_2 (L-t)\right]\right\}
\label{eq:corrfun}
\end{eqnarray}
where $m_1$ is the mass of the $\rho$ meson, while $m_2$ is the mass of its opposite
parity partner, whose contribution has an alternating sign
in the time direction.
The partner for $\rho(1)$ is the $b_1$ meson, with spin-taste 
$(\gamma_j\gamma_k \otimes \xi_j\xi_k)$, while that for the $\rho(2)$ is 
the $a_1$ with spin-taste $(\gamma_i\gamma_5 \otimes \xi_i\xi_5)$. 
We obtain good fits to both $\rho$ correlators except 
for the $\rho(1)$ at the lightest quark mass. 
The resulting masses, as well as those of
the parity partner $a_1$, are given in Table \ref{tab:rho_mass}.
Since we want to be able
to use all four quark masses to carry out the chiral extrapolation, we opt
to consider only the $\rho(2)$ results to determine the lattice scale.
We do not, however, expect that the resulting scale would change significantly
were we to use $\rho(1)$ masses, because the $\rho(1)$ and $\rho(2)$ masses
agree within errors for the three heavier quark masses.
To illustrate the quality of the fits we show 
the effective mass plots for $\rho(2)$ as a function of time 
in Figs.~\ref{fig:mrho_m1}-\ref{fig:mrho_m4}.
The effective mass at time $t=T$ is defined to be the value of $m_1$ obtained
by solving Eq.~(\ref{eq:corrfun}) 
using the correlation function on time slices $T$ to $T+3$.
All errors are determined using single elimination jackknife, with the
underlying fits (both in $t$ and $m$) using uncorrelated errors, 
because the correlation matrix is determined with insufficient
accuracy.

\begin{figure}[t!]
\includegraphics[width=20pc]{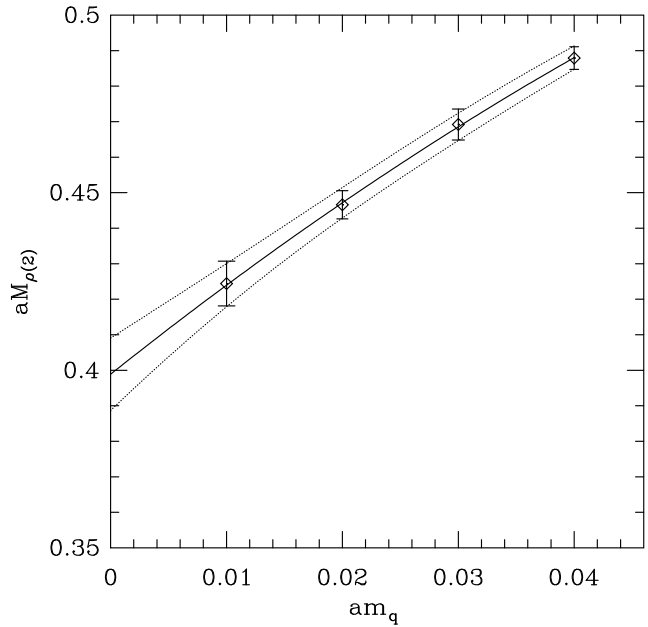}
%%%\vspace*{-5mm}
\caption{$aM_{\rho}$ vs. quark mass.
}
\label{fig:mrho_mq}
%%%\vspace*{-3mm}
\end{figure}

The result of a quadratic fit to $M_{\rho(2)}$ versus quark mass, 
shown in Fig.~\ref{fig:mrho_mq}, is 
\begin{equation}
a M_\rho = 0.399(10) + 2.60(59) (a m_q) - 9.3(8.7) (a m_q)^2 \,.
\end{equation}
From this we estimate the lattice scale $1/a$ quoted in Table
\ref{tab:rho_mass} by setting the chirally extrapolated value
$0.399(10) = a M_\rho^{physical}$. The change in the resulting scale from
extrapolating to the physical light quark masses 
($a (m_u+m_d)/2 \sim 0.0015$) rather than the chiral limit
is smaller than our statistical errors and we do not include it.
We also do not include the $m_q^{1/2}$ and $m_q^{3/2}$ terms
from pion loops~\cite{rhochipt}
in our chiral fit, as they are expected to have small coefficients, and
we have too few mass points to reliably include them.

A potential problem with our estimate of the scale is our use of
a relatively small volume ($L\approx 1.6\;$fm). Although we expect
this is large enough to study kaon properties (since $m_K L\approx 5$ is larger
than the range $3-4$ where significant effects usually set in),
we are relying in our scale determination on results from all four quark
masses. At the lightest quark mass $M_\pi L=2.7$, and volume errors
may be significant. Evidence that this is the case comes from Ref.~\cite{ref:aoki:1},
who have results for $M_\rho$ for three different volumes at $\beta=6$.
The finite volume effects can be seen
from the resulting estimates of the scale:
$1/a=1.87(6)$, $1.88(4)$ and $2.01(2)\,$GeV, 
for $18^3$, $24^3$ and $32^3$ lattices, respectively.
Thus we may have underestimated the scale by $\approx 7\%$.
It turns out, however, that this uncertainty is smaller than the
range of scales resulting from the use of different physical quantities,
and so can be subsumed into the quenching error discussed below.

\begin{figure}[t!]
\includegraphics[width=20pc]{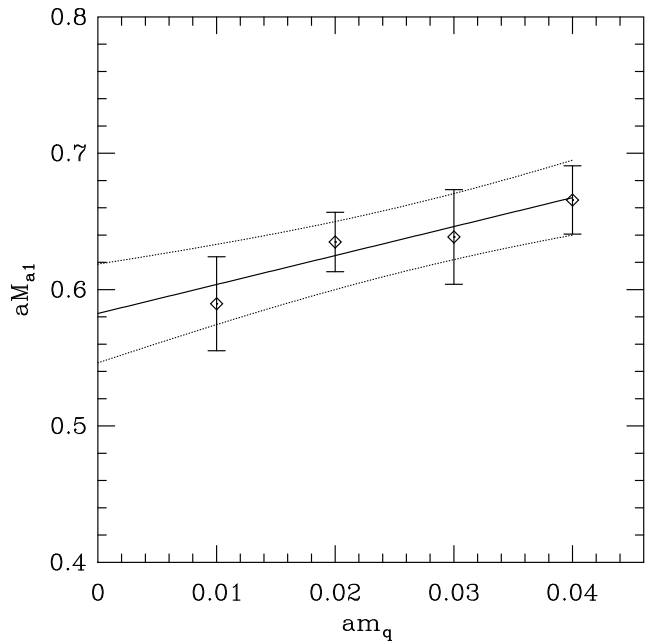}
%%%\vspace*{-5mm}
\caption{$aM_{a1}$ vs. quark mass.}
\label{fig:ma1_mq}
%%%\vspace*{-3mm}
\end{figure}

Data for the $a_1$ meson has much larger errors and we use a simple
linear fit. The result,
shown in Fig.~\ref{fig:ma1_mq}, gives 
\begin{equation}
a M_{a_1} = 0.58(4) + 2.1(1.0) (a m_q)
\end{equation}
Again, the chirally extrapolated
value is used to determine the estimate of $1/a$ given in Table \ref{tab:rho_mass}.
%

%% estimates from Weonjong on 01/24/05
%% For rho(2), c_0 = 0.3989(102)  
%% 	    c_1 = 2.5989(5908)
%% 	    c_2 = -9.2543(87468)
%% 	    chi2/dof = 0.0507
%% 
%% For a_1,    c_0 = 0.5825(362)
%% 	    c_1 = 2.1240(10118)
%% 	    chi2/dof = 0.2150
%% 
%% For M_pi2, c_0 = 0.0045(9)
%% 	    c_1 = 2.3190(308)
%% 	    chi2/dof = 0.1433(947)
%% 

One of the well-known uncertainties introduced by quenching is that
different physical quantities lead to different values of the lattice
spacing. Since we are interested here in comparing with other quenched
results for $m_s$ and $B_K$, we follow most previous calculations and
determine our central value for the scale, $1/a=1.95$ GeV, using
$\rho$ masses.  This lies within the range of values quoted above
from the JLQCD $B_K$ calculation, and is close to the value,
$1/a=1.855(38)\,$GeV, they use when estimating $m_s$~\cite{ref:aoki:0}.  
Nevertheless, to understand
the impact of the scale uncertainty, and, as noted above,
to include possible finite volume errors, we also analyze subsequent data 
using $1/a=2.1$ GeV.
This value is consistent with our estimate from $a_1$ as well as 
the result obtained
using the Sommer parameter $r_0$~\cite{ref:sommer} ($1/a=2.12\,$GeV).
The latter is derived from the
static $q \overline q$ potential and thus is independent of the fermion
action.

%

% {\bf Figure 1-4: x and y-axis label should match text --- $M_{\rho(2)}$.
% Caption should also mention "effective mass".
% Similarly in Figs 5-7 $aM_q$ should be replaced by $am_q$.}

%

%

%%  \begin{figure}[t!]
%%  \includegraphics[width=20pc]{a1.m1.eps}
%%  %%%\vspace*{-5mm}
%%  \caption{$m_{a1}$ at quark mass 0.01.
%%  }
%%  \label{fig:ma1_m1}
%%  %%%\vspace*{-3mm}
%%  \end{figure}
%%  %
%%  
%%  \begin{figure}[t!]
%%  \includegraphics[width=20pc]{a1.m2.eps}
%%  %%%\vspace*{-5mm}
%%  \caption{$m_{a1}$ at quark mass 0.02.
%%  }
%%  \label{fig:ma1_m2}
%%  %%%\vspace*{-3mm}
%%  \end{figure}
%%  %
%%  
%%  \begin{figure}[t!]
%%  \includegraphics[width=20pc]{a1.m3.eps}
%%  %%%\vspace*{-5mm}
%%  \caption{$m_{a1}$ at quark mass 0.03.
%%  }
%%  \label{fig:ma1_m3}
%%  %%%\vspace*{-3mm}
%%  \end{figure}
%%  %
%%  
%%  \begin{figure}[t!]
%%  \includegraphics[width=20pc]{a1.m4.eps}
%%  %%%\vspace*{-5mm}
%%  \caption{$m_{a1}$ at quark mass 0.04.
%%  }
%%  \label{fig:ma1_m4}
%%  %%%\vspace*{-3mm}
%%  \end{figure}
%

\begin{table}[t!]
\caption{Masses of $\rho$ and $a_1$ mesons, and resulting scales.
\label{tab:rho_mass}}
\begin{ruledtabular}
\begin{tabular}{c | l l l}
  $m_q$ &  $a M_\rho(\gamma_i \otimes \xi_i)$ & 
           $a M_\rho(\gamma_i\gamma_4 \otimes \xi_i\xi_4)$ & 
           $a M_{a_1}(\gamma_i\gamma_5 \otimes \xi_i\xi_5)$ \\ 
\hline
  $m_1$ & $-$        & 0.4244(63) & 0.5897(345) \\
  $m_2$ & 0.4444(32) & 0.4466(40) & 0.6350(218) \\
  $m_3$ & 0.4676(32) & 0.4692(43) & 0.6387(347) \\
  $m_4$ & 0.4865(25) & 0.4879(32) & 0.6657(250) \\
\hline
  $1/a$ & $-$ & 1945(50) MeV & 2112(131) MeV  
\end{tabular}
\end{ruledtabular}
\end{table}
%
%---------------------------
% 1/a = 1.95
%---------------------------

\section{Strange quark mass}
\label{sec:quark mass}

Our results for the masses of the lattice psuedo-Goldstone pion
(spin-taste $\gamma_5 \otimes \xi_5$) are presented in
Table~\ref{tab:pion_mass}. The results with different sources and
sinks are consistent, and we use the weighted average of the four
results in the subsequent analysis.

\begin{table}[t!]
\caption{ Pion masses using axial and pseudoscalar 
operators from Left ($t=10$) and Right ($t=36$) wall sources.}
\label{tab:pion_mass}
\begin{ruledtabular}
\begin{tabular}{c | l l l l}
  $m_q$ &  $a M_\pi(A_4,L)$ & $a M_\pi(A_4,R)$ & 
  $a M_\pi(P,L)$ & $a M_\pi(P,R)$ \\ 
\hline
  $m_1$ & 0.1697(29) & 0.1682(30) & 0.1658(50) & 0.1644(56) \\
  $m_2$ & 0.2266(27) & 0.2255(28) & 0.2248(40) & 0.2224(42) \\
  $m_3$ & 0.2732(27) & 0.2725(26) & 0.2716(35) & 0.2695(34) \\
  $m_4$ & 0.3136(27) & 0.3134(24) & 0.3120(32) & 0.3106(30)
\end{tabular}
\end{ruledtabular}
\end{table}
%

\begin{figure}[t!]
\includegraphics[width=20pc]{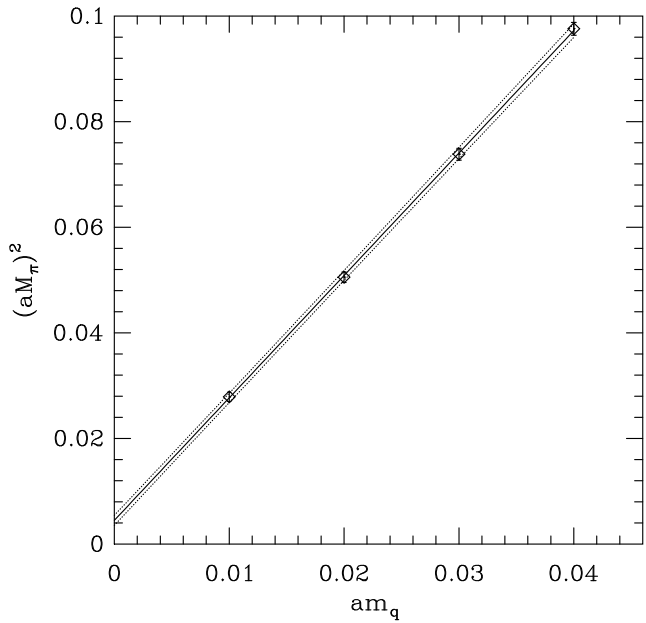}
%%%\vspace*{-5mm}
\caption{$(aM_{\pi})^2$ vs. quark mass.}
\label{fig:mpi2_mq}
%%%\vspace*{-3mm}
\end{figure}

%%%%%%%%%%%%%%%%%%%%%%%%%%%%%%%%%%%%%%%%
% m_s
%%%%%%%%%%%%%%%%%%%%%%%%%%%%%%%%%%%%%%%%
%
The strange quark mass $m_s$ is determined by requiring a fictitious
$\bar ss$ pseudoscalar mass to match the physical value of $(2 M_K^2 -
M_\pi^2)$ which corresponds to $(aM_{PS})^2= 0.1234$ with
$1/a=1.95\,$GeV.  A linear fit for $(aM_{PS})^2$ versus $a m_q$ works
well, as shown in Fig.~\ref{fig:mpi2_mq}. This fit gives $a m_s =
0.0520(7)$, and thus $m_s = 102(1.3)$ MeV. Repeating the analysis with
$1/a=2.1\,$GeV leads instead to $110(1.5)\,$MeV. We stress that these
results are not very sensitive to the chiral fit form used.  For
example, a fit that includes a quenched chiral 
logarithm~\cite{BG,ref:sharpe:1} and is forced
to pass through the origin reduces $m_s$ by 2.1(6) MeV.
Such consistency is not surprising since $a m_s = 0.052$ is larger
than the simulated points, whereas quenched chiral logarithms are
important only at masses smaller than $am_q = 0.01$.

As discussed in the previous section, we expect that finite volume
effects should be small in our determination of $m_s$, because our heaviest
two quark masses dominate the determination, and these have 
relatively large values of $M_\pi L$, $4.3$ and $5.0$ respectively.
According to the quenched chiral perturbation theory analysis of Ref.~\cite{ref:sharpe:1},
one would expect $M_\pi^2$ to be larger than its infinite volume value
by  1-2\% in this quark mass range. This would lead to our finite volume result
for $m_s$ being 1-2\% smaller than the infinite volume value.
This estimate assumes the quenched hairpin parameter to be $\delta\approx 0.2$;
using more recent values of $\delta\approx 0.1$ reduces the effect proportionally.
That the finite size effects are no larger than this size is supported by the numerical analysis of finite volume effects given in Ref.~\cite{aokifv}.

Finite volume errors also enter into our result for $m_s$ through their
effect on the scale, as discussed in the previous section. We choose,
however, to quote a value for $m_s$ for a definite choice of scale,
so as to allow more straightforward comparison with other results.
In particular, using the one-loop matching factor from Ref.~\cite{ref:wlee:6},
and the scale $1/a=1.95\,$GeV, we find the renormalized mass
$m_s(\overline{\rm MS},2\; {\rm GeV}) = 101.2 \pm 1.3 \pm 4\;$MeV.
Here the first error is statistical, while the second is from the
systematic effects that we control aside from the scale uncertainty. It is dominated by 
the uncertainty in $Z_m$, which we estimate as $4\%$ by 
assuming a two-loop term of size
$\pm 1\times (\alpha_s)^2$.  It also contains the uncertainty from the form of
chiral fit used, and from the finite volume errors in $M_\pi^2$ discussed in
the previous paragraph.
Note that we take
the central value from the fit form without quenched chiral logarithms
so as to better compare to the results of Ref.~\cite{ref:aoki:0}.

This result for $m_s$ allows us to study the efficacy of HYP improved
staggered fermions.  The state-of-the-art quenched estimate for
unimproved staggered quarks (obtained using the same definition of
$m_s$, and $M_{\rho(1)}$ for setting the scale) is $m_s(\overline{\rm
MS},2{\rm GeV}) = 106.0 \pm7.1\;$MeV, after extrapolation to the
continuum limit~\cite{ref:aoki:0}.  Our first observation is that our
result at $\beta=6.0$ agrees with this continuum value, consistent
with our expectation that $a^2$ errors should not be large.  In this
respect, we note that it was necessary to go down to lattice spacing
$a=0.06\,$fm with unimproved staggered fermions in order to obtain the
continuum estimate~\cite{ref:aoki:0}.

It is also useful to compare with the results from
Ref.~\cite{ref:aoki:0} obtained at our coupling, $\beta=6$. Their bare
quark mass, $am_s=0.0244$ or $m_s = 45\;$MeV, is much smaller than ours.
Using non-perturbative renormalization they find $m_s(\overline{\rm
MS},2{\rm GeV}) = 114\,$MeV.  The very large matching factor,
$Z_m\approx 2.5$, shows the need for non-perturbative renormalization
with unimproved staggered fermions. Indeed, using one-loop matching
they find the significantly smaller value $m_s(\overline{\rm MS},2{\rm
GeV})=84\,$MeV.  By contrast, our matching factor is very close to
unity, illustrating one of the advantages of HYP smeared staggered
fermions.

Quantifying improvement in discretization errors is more difficult.
The unimproved (but non-perturbatively renormalized) result drops by
$8\,$MeV between $\beta=6$ and the continuum, whereas our result is
$5\,$MeV lower than the continuum value. Since these differences are
comparable to the errors, the only definite conclusion we can draw is
that the discretization errors appear to not be worsened by
improvement.\footnote{This conclusion is bolstered by the fact that
Ref.~\cite{ref:aoki:0} uses a smaller scale than us ($1.85\,$ rather
than $1.95$\,GeV). Had they used the larger scale, their final result
at $\beta=6$ would have differed more from the continuum value.}

\section{$B_K$}
\label{sec:B_K}

\begin{figure}[t!]
\includegraphics[width=20pc]{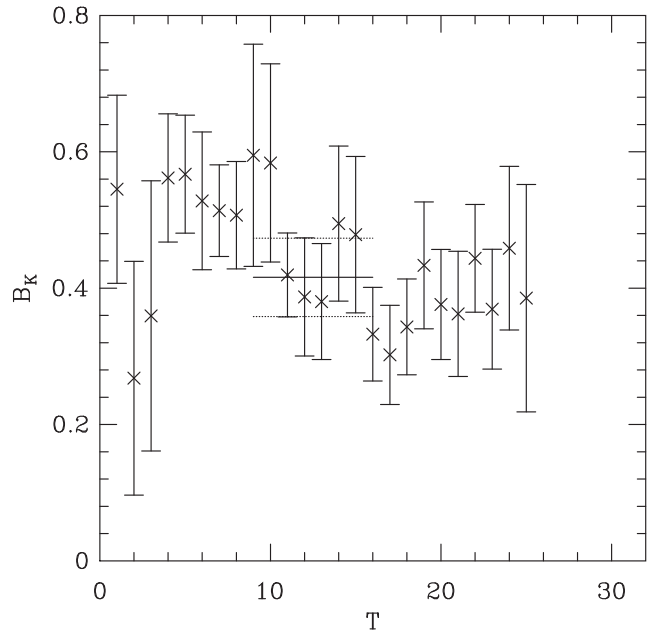}
%%%\vspace*{-5mm}
\caption{$B_K$ at quark mass 0.01.
}
\label{fig:bk_m1}
%%%\vspace*{-3mm}
\end{figure}
\begin{figure}[t!]
\includegraphics[width=20pc]{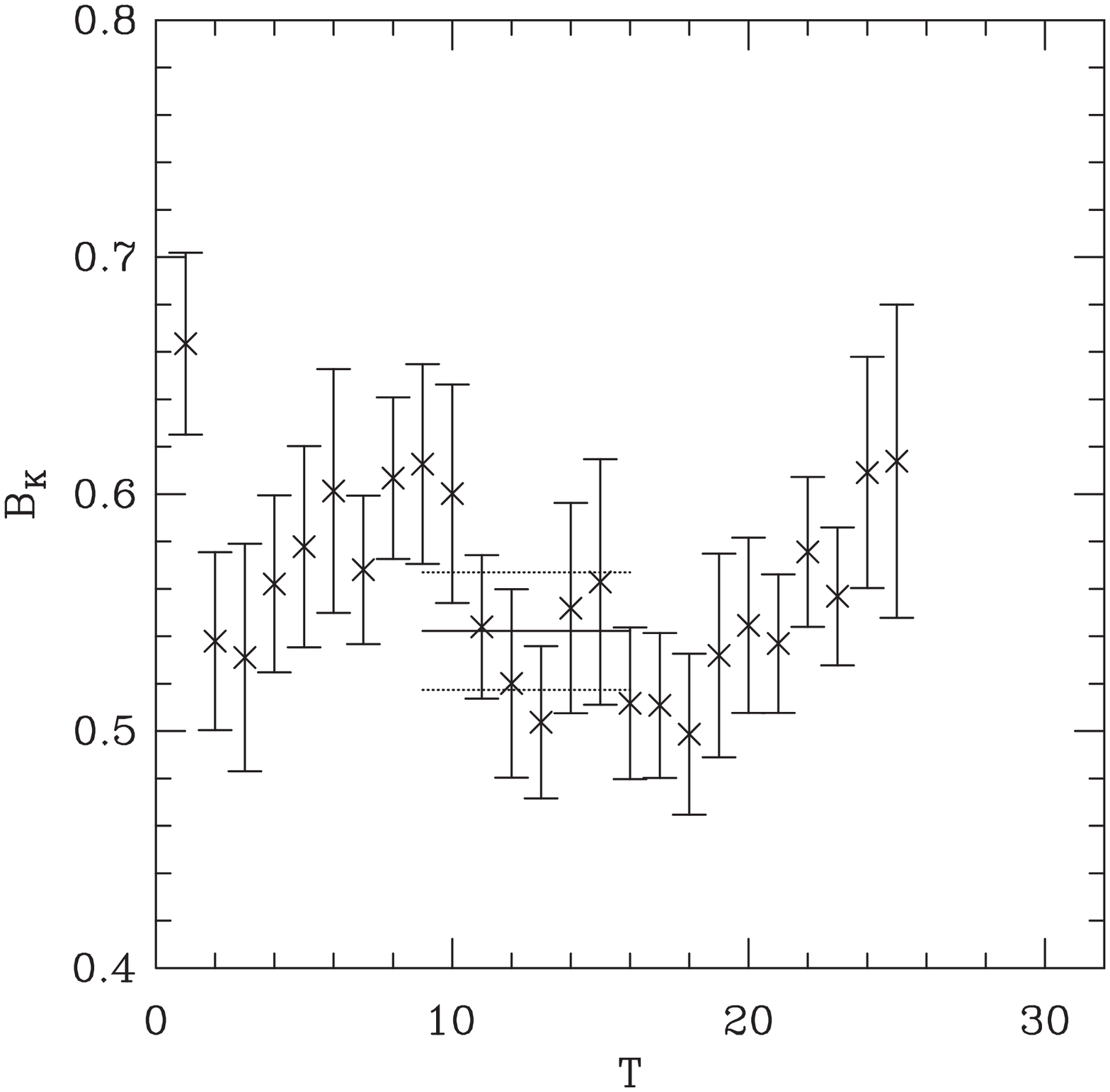}
%%%\vspace*{-5mm}
\caption{$B_K$ at quark mass 0.02.
}
\label{fig:bk_m2}
%%%\vspace*{-3mm}
\end{figure}
\begin{figure}[t!]
\includegraphics[width=20pc]{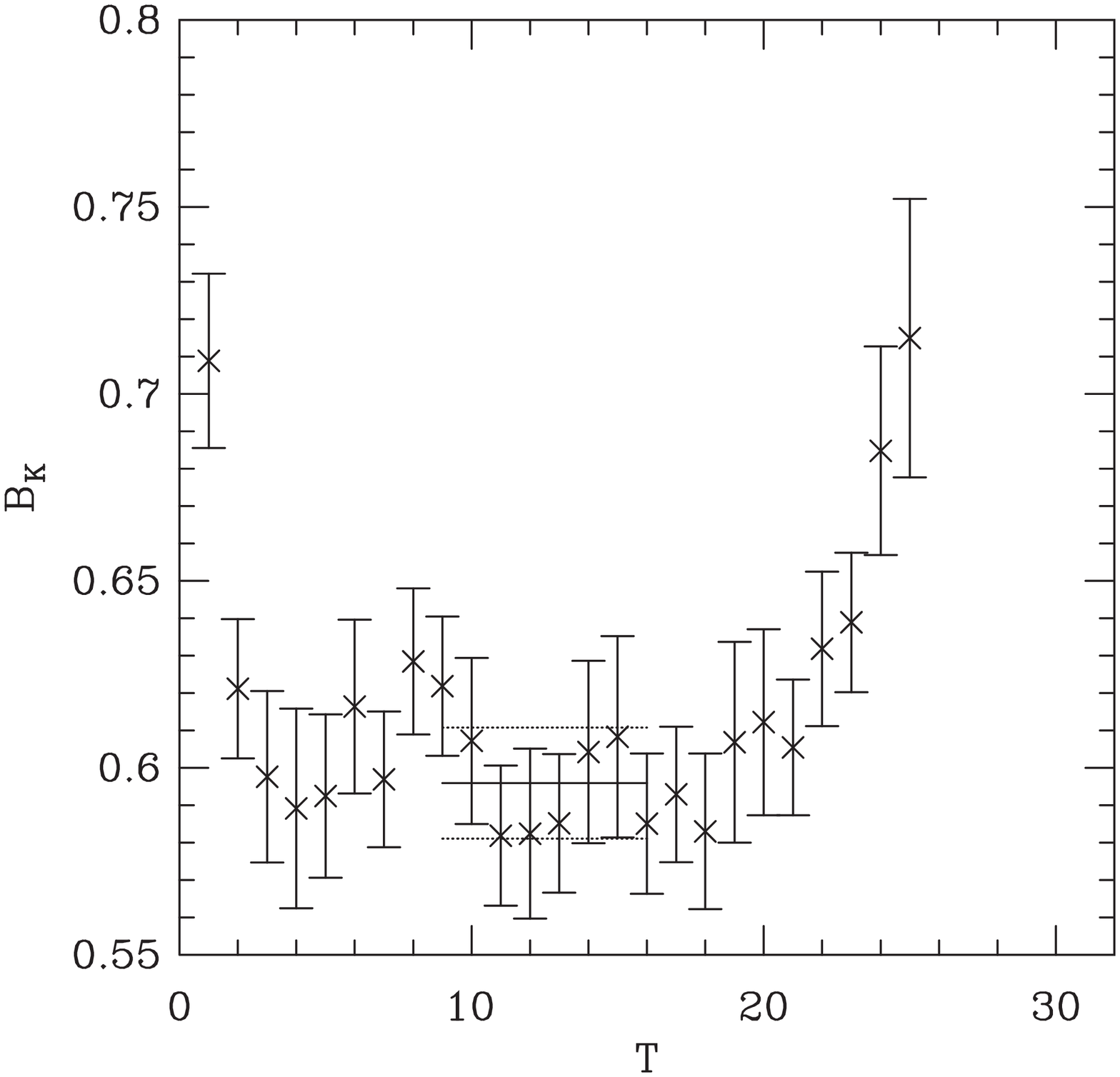}
%%%\vspace*{-5mm}
\caption{$B_K$ at quark mass 0.03.
}
\label{fig:bk_m3}
%%%\vspace*{-3mm}
\end{figure}
\begin{figure}[t!]
\includegraphics[width=20pc]{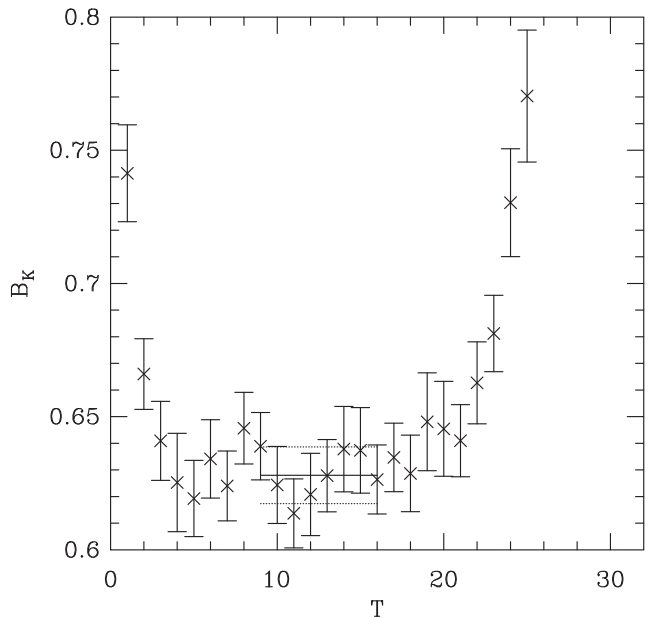}
%%%\vspace*{-5mm}
\caption{$B_K$ at quark mass 0.04.
}
\label{fig:bk_m4}
%%%\vspace*{-3mm}
\end{figure}
The ratio of correlators corresponding to Eq.~\ref{eq:bkdef} is
measured on the interval $ 1 \le t \le 25 $ between two random
$U(1)$ sources~\cite{u1sources}
 placed at $t=0$ and $26$. We find that the individual
pseudoscalar meson correlators exhibit contamination from excited
states up to $\approx 9$ time slices from the sources.  For this
reason we choose to make constant fits to the central part of the
plateau on time slices $ 9 \le t \le 16 $ even though the estimate is
stable over the range $ 4 \le t \le 20$. These fits are shown in
Figs.~\ref{fig:bk_m1}-\ref{fig:bk_m4}.  In the first column of
Table~\ref{tab:BK} we give the resulting bare values for $B_K$, $i.e.$
with all renormalization constants set to unity, for each of the quark
masses. In the second column we give the results after renormalization
to $\overline{\rm MS}, {\rm NDR}$ scheme at scale $\mu = 1/a$.

To quote results in the $\overline{\rm MS}$ scheme we use the one loop
renormalization factors of Ref.~\cite{ref:wlee:4}, with the matching
scale chosen to be $q^*=1/a$.  The coupling $\alpha_s(q^*=1/a)=0.192$
is calculated from the plaquette expectation value ($P=0.59367$) using
the method of Ref.~\cite{ref:davies:1}.
%
%--------------------------------------------------------------
%\footnote{%
%To be precise, $\alpha_P(3.41/a)$ is obtained from $P$
%at three loop order,
%run to the scale $e^{(5/6)} 3.41/a$, 
%converted to $\alpha_{\overline{MS}}(3.41/a)$
%at two-loop accuracy,
%and finally evolved to the needed scale at three loop.
%{\bf It would be better to skip the first running, but not essential.}}
%
%$g^2_{\overline{\rm MS}}(1/a)=2.4142721$.
%--------------------------------------------------------------
%
At the physical kaon mass, which corresponds to $am_q =0.026$, the
one-loop corrections lead to a $\sim 10\%$ change in $B_K$.  This is
very similar to the corresponding shift with unimproved staggered
fermions.

\begin{table}[t!]
\caption{Results for bare $B_K$, $B_K({\rm NDR},\mu=1/a,L)$ and
estimate of finite volume shift: $\delta B_K(L)=B_K(L=\infty)-B_K(L)$.
Errors are statistical.}
\label{tab:BK}
\begin{ruledtabular}
\begin{tabular}{c | l l l}
$m_q$ & bare $B_K$ & $B_K(1/a,L)$ & $\delta B_K(1/a,L)$ \\
\hline
$m_1$ & 0.514(61) & 0.416(57) & $+0.0094(38)$ \\
$m_2$ & 0.614(26) & 0.542(25) & $+0.0013(5)$  \\
$m_3$ & 0.658(16) & 0.596(15) & $-0.0003(1)$  \\
$m_4$ & 0.686(11) & 0.628(11) & $-0.0006(2)$  
\end{tabular}
%\begin{tabular}{c | l l l}
%$m_q$ & $B_K(1/a,L)$ & $\delta B_K(1/a,L)$ & $B_K(1/a,L=\infty)$ \\
%\hline
%$m_1$ & 0.416(57) & $-0.0101(43)$ & 0.440(54) \\
%$m_2$ & 0.542(25) & $-0.0014(6)$  & 0.537(27) \\
%$m_3$ & 0.596(15) & $+0.0003(1)$  & 0.598(15) \\
%$m_4$ & 0.628(11) & $+0.0006(2)$  & 0.627(11)
%\end{tabular}
\end{ruledtabular}
\end{table}
%

%
%--------------------------------------------------------
% finite volume correction
% a = hbar * c / (1/a) = hbar * c / 1.95 GeV = 0.101 fm
%--------------------------------------------------------
%
%
\begin{figure}[t!]
\includegraphics[width=20pc]{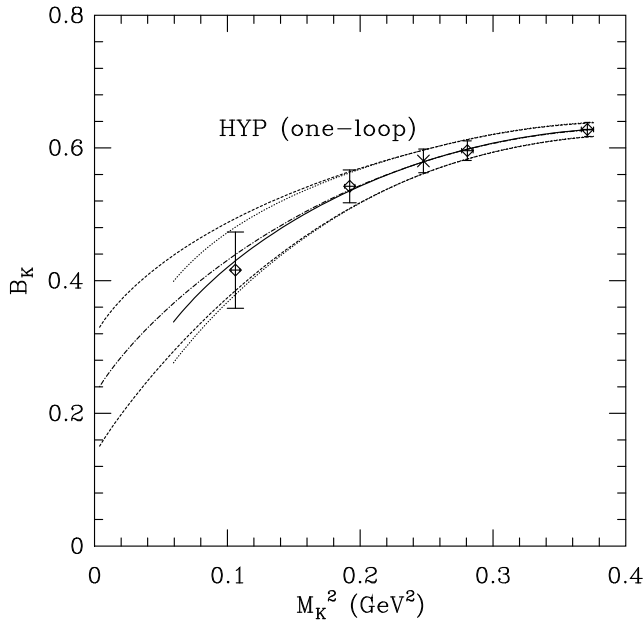}
%%%\vspace*{-5mm}
\caption{Results for $B_K(NDR,\mu=1/a)$ (diamonds), fit to the
expected chiral form at finite volume (solid line, errors small dashes).
The dot-dashed line (errors long dashes)
is the corresponding infinite volume result.
The cross is the infinite volume result at the physical kaon mass.
%Note that the ``kaon'' is composed of degenerate quarks.
}
\label{fig:BK}
%%%\vspace*{-3mm}
\end{figure}

To extract $B_K$ at the physical kaon mass 
we fit the data to the form predicted by quenched
chiral perturbation theory including 
finite volume corrections~\cite{ref:sharpe:1}
\begin{eqnarray}
B_K &=& b_0 \Bigg\{1 - 6.0 \frac{M_K^2}{ (4 \pi f)^2 }
\log\left[\frac{M_K^2}{ (4 \pi f)^2 }\right]
\nonumber \\
% Damir's notation
%+ \frac{1}{2 f^2} \left[ - 4 \xi_{1/2}(L,M_K) +  M_K^2 \xi_{3/2}(L,M_K) \right]
% G+L notation
&& + \frac{2}{ f^2} \left[ - 2 g_1(M_K^2,0,L) +  M_K^2 g_2(M_K^2,0,L) \right]
\Bigg\}
\nonumber \\
&& + b_1  M_K^2 + b_2  M_K^4 
\label{eq:bk:fvol}
\end{eqnarray}
where $f$ is the decay constant, which we fix to $132\;$MeV.
The finite volume dependence enters through the functions
$g_i$, defined in Ref.~\cite{ref:GLfiniteL}.
%\footnote{%
%In the more recent notation of Ref.~\cite{ref:becirevic:1} the
%finite volume part is
%$[1/(2 f^2)] [ - 4 \xi_{1/2}(L,M_K) +  M_K^2 \xi_{3/2}(L,M_K)]$.}
%
%We evaluate these functions to sufficient accuracy
%using the method explained in~\cite{ref:sharpe:1}.
%
This dependence of $B_K({\rm NDR},\mu=1/a)$ on the quark mass is shown
by the solid line in Fig.~\ref{fig:BK} with parameter values
$b_0=0.23(9)$, $b_1=0.5(1.1)/{\rm GeV}^2$ and $b_2=-1.2(1.3)/{\rm GeV}^4$.
%
%\begin{eqnarray}
%b_0 &=&  0.231 \pm 0.090 \,, \\
%b_1 &=&  0.48  \pm 1.12\ {\rm GeV}^{-2} \,, \\
%b_2 &=&  -1.1  \pm 1.4 \ {\rm GeV}^{-4} \,.
%\label{eq:fit:b:fvol}
%\end{eqnarray}
% $b_0^L =  0.2308 \pm 0.0897$ at $\mu=2GeV$ in the NDR scheme.
% $b_1^L =  0.4809 \pm 1.1218$ at $\mu=2GeV$ in the NDR scheme.
% $b_2^L = -1.1332 \pm 1.3549$ at $\mu=2GeV$ in the NDR scheme.
%
Since the prediction for the finite volume corrections becomes
unreliable once $M_K L$ becomes small, we do not display the fit
function below $M_K L=2$.

Our results are consistent with the curvature predicted by the chiral
logarithm in Eq.~(\ref{eq:bk:fvol}).  Indeed, we can set $b_2=0$ and
obtain a good fit [with $b_0=0.283(29)$ and $b_1=-0.30(19)$
GeV${}^{-2}$], showing that the curvature can be accounted for by the
logarithm alone.  Another consistency check is that $b_1$ and $b_2$
agree with the expectations of naive dimensional analysis, namely
$|b_1| \approx |b_2| \approx 1$ in units of the scale, $\sim 1$ GeV,
of chiral perturbation theory.  Taken as a whole, previous work is
inconclusive concerning the presence of the chiral logarithm with
predicted coefficient, largely due to the relatively high quark masses
used ($m > m_s/2$).  It is only by extending the range to $m_s/5$ that
we find evidence, albeit not conclusive, for the onset of the expected
chiral logarithm at small quark masses.

It is important to obtain a good fit to the chiral behavior in order
to reliably extract estimates of finite volume corrections. The
smallest value of $M_K L$ is 2.64, so one expects such corrections to
be large~\cite{ref:becirevic:1}. For $B_K$, however, the cancellation
between $g_1$ and $g_2$ terms suppresses these corrections.  Based on
our chiral fit, the third column in Table~\ref{tab:BK} gives estimates
for this finite volume shift. Note that the correction is
non-monotonic in $M_K$, due to the cancellation noted above.  The
infinite volume prediction is shown in Fig.~\ref{fig:BK}.  From this
we conclude that for physical $M_K$ the finite volume shift in $B_K$
is much smaller than quoted errors even on our small lattices.
This conclusion is supported by the absence of finite volume errors
in the JLQCD results
 at $\beta=6$ using unimproved staggered fermions~\cite{ref:aoki:1}.

Our final results are obtained by evolving from $1/a$ to $\mu=2\;$GeV
using two-loop renormalization group running for
$N_f=0$~\cite{ref:buras:1}.  We find
%
%\begin{eqnarray}
%B_K^{NDR}(\mu) &=& 
%\frac{ [ 1 - \frac{\alpha(q^*)}{4\pi } Z ] }
%{ [ 1 - \frac{\alpha(\mu)}{4\pi} Z ] }
%\bigg( \frac{\alpha(\mu)}{\alpha(q^*)} \bigg)^{ d^{(0)} } 
%B_K^{NDR}(q^*) \nonumber 
%\label{eq:b_k:1}
%\\
%Z &=& \frac{ \gamma^{(1)} }{2 \beta_0} - 
%d^{(0)} \frac{ \beta_1 }{ \beta_0 }  \\
%d^{(0)} &=& \frac{ \gamma^{(0)} }{ 2 \beta_0 }
%\nonumber
%\end{eqnarray}
%
%where $\alpha$ is the gauge coupling
%renormalized in the $\overline{MS}$ scheme, and
%%
%$\beta_0 = 11$, $\beta_1 = 102$, $\gamma_0=4$ and $\gamma_1 = -7$ are
%RG coefficients for $N_f=0$.
%%
%
%
\begin{eqnarray}
B_K({\rm NDR},2\,{\rm GeV}) &=& 0.578 \pm 0.018 \pm 0.042 \,, 
\label{eq:BKNDRres}\\
B_K^{RGI}        &=& 0.806 \pm 0.025 \pm 0.058 \,, \\
b_0^{RGI}        &=& 0.314 \pm 0.124 \pm 0.176 \,.
\label{eq:bk:infvol}
\end{eqnarray}
where $B_K^{RGI}$ is the renormalization group invariant
$B-$parameter~\cite{ref:buras:1}, with $b_0^{RGI}$ its value in the
chiral limit. The first error combines that from statistics and those
due to the chiral interpolation (or extrapolation for $b_0$).  The
second is our estimate of the uncertainty from using one-loop matching
factors explained below.
Aside from the errors due to quenching and the use of degenerate
quarks, which we do not address here, other systematics lead to
changes smaller than the perturbative error.  For example, using the
scale from $r_0$ reduces $B_K({\rm NDR},2\,{\rm GeV})$ and $b_0^{RGI}$
by $0.025$ and $0.007$, respectively, while setting $b_2=0$ in the
chiral fit increases them by $0.011$ and $0.079$.
If we use $f = f_K = 159.8$ MeV instead of $f=132$MeV in
Eq.~\ref{eq:bk:fvol} and fit the data, $B_K$ changes by less than
0.01\%, while $b_0$ increases by 5\%.

The error associated with unknown $\alpha^2$ corrections is estimated
as follows.  We can write $B_K= B_{A} + B_{V}$, where $V$ and $A$
refer to vector-vector and axial-axial parts of the operator in
Eq.~(\ref{eq:bkdef}). $B_{A,V}$ can each be decomposed into one and
two color-trace parts~\cite{ref:smearedops}.  Each of these four
components of $B_K$ is proportional to $\log(M_K^2)$ and thus diverges
in the chiral limit, although their sum does not.  Using one-loop
matching there is an incomplete cancellation, and the resulting $B_K$
should diverge in the chiral limit, although this feature is expected
to manifest itself at much smaller quark masses than studied here.
Indeed, the bare values for $B_{V,A}$
%, shown in Fig.~\ref{fig:b_k:3},
do indicate a divergent behavior.  Because of the residual divergence,
we cannot simply estimate the error, in particular in $b_0$, by
multiplying by an overall relative correction of $\pm \alpha(q^*)^2$
(as we did for $m_s$).
%, but there is no sign of it in $B_K$ itself. 
Instead, we recalculate $B_K$ after adding $\pm \alpha(q^*)^2$ to the
matching factors for each of the four components of $B_K$ in turn, and
take the largest variation as the error.  The resulting uncertainty,
quoted above, is larger than the statistical error and, as expected,
grows rapidly in the chiral limit.
%
%{\bf Weonjong: so we don't change more than one of the four Z's at a time?}
%

Even though we need more high precision data at lighter quark masses
to pin down the chiral extrapolation, it is nevertheless interesting
that our estimate $b_0^{RGI} = 0.314 \pm 0.124 \pm 0.176 $ is in good
agreement with recent estimates $0.29(15) $~\cite{Prades:BK:04} and
$0.36(15)$~\cite{ref:peris:1} obtained using $1/N_c$ expansion.

%
%\begin{figure}[t!]
%\epsfig{file=../figs/BA_BV.eps, height=16pc, width=16pc}
%%%%\vspace*{-5mm}
%\caption{Bare values of $B_A$ and $B_V$.}
%\label{fig:b_k:3}
%%%%\vspace*{-3mm}
%\end{figure}
%

We now compare our estimate with the state-of-the-art results obtained
by the JLQCD collaboration~\cite{ref:aoki:1} using unimproved
staggered fermions and argue that HYP smearing reduces discretization
errors.  The JLQCD result in the continuum limit is $B_K({\rm
NDR},2\;{\rm GeV})=0.628 \pm 0.042$.  Our first observation is that
our result at $\beta=6$ is consistent with this continuum result. On
the one hand, this indicates that the $a^2$ errors with HYP fermions
are not large, as for $m_s$.  On the other hand, the value of $B_K$ at
$\beta=6$ with unimproved staggered fermions [$0.679(2)$] is also
consistent with the continuum result, giving no evidence of
improvement.

We can go further, however, using the details of the continuum
extrapolation provided by Ref.~\cite{ref:aoki:1}. Their fit included
both $a^2$ and $\alpha_s^2$ terms.  Using their fit parameters we can
determine two additional estimates of $B_K$ at $\beta=6.0$: removing
only the $O(\alpha^2)$ term (and not the $O(a^2)$ discretization
correction) and vice-versa.
The original results and those corrected for either the discretization
or perturbative errors alone are given in Table~\ref{tab:jlqcd}.  The
point we wish to make is that in the case of gauge invariant operators
(which are those we use) the JLQCD fits imply that, at $\beta=6$, the
total result $0.68$ contains an $O(a^2)$ contribution of $\sim 0.13$
and an $O(\alpha^2)$ contribution of $\sim -0.08$.  (Corrections for
non gauge-invariant operators, which are not the operators of choice,
are somewhat smaller but show the same pattern.) Thus, in a
formulation where only discretization errors were eliminated or
substantially reduced one should expect a final result closer to
$0.55$ at $a\approx 0.1\;$ fermi.  Our estimate with HYP smearing,
$B_K({\rm NDR},2\;{\rm GeV})=0.58(4)$, is indeed consistent with this
and significantly different from the unimproved JLQCD result
$0.679(2)$.  This suggests that discretization errors have been
reduced by using HYP smeared fermions.

\begin{table}[t!]
\caption{$B_K({\rm NDR},2\;{\rm GeV})$
at $\beta=6$ for gauge invariant (GI) and 
non-invariant (NGI) operators, before and after
removing the fitted $a^2$ and $\alpha^2$ terms.
Data from Ref.~\cite{ref:aoki:1}.}
\label{tab:jlqcd}
\begin{ruledtabular}
\begin{tabular}{c | l l l}
Type &  Uncorrected  & $a^2$ removed & $\alpha^2$ removed \\
\hline
GI  & 0.6790(16) & 0.55(7)  & 0.76(7)\\
NGI & 0.7128(14) & 0.61(7)  & 0.73(7)
\end{tabular}
\end{ruledtabular}
\end{table}

This conclusion is supported by the fact that all calculations with
domain wall or overlap fermions, which are also expected to have small
discretization errors and small perturbative corrections, find values
at $\beta=6$ consistent with ours: 0.575(6)
(Ref.~\cite{ref:cp-pacs:1}), 0.532(11) (Ref.~\cite{ref:rbc:1}),
$0.563(31)$ (mean of data at $\beta=5.9$ and $6.1$ in
Ref.~\cite{ref:degrand:1}), and 0.63(6) (Ref.~\cite{ref:lellouch:1}).
Here, only statistical errors have been quoted.

Finally, we consider the results in Table~\ref{tab:jlqcd} with
``$\alpha^2$ removed". These correspond approximately to using
non-perturbative matching factors, and should thus expose the ``true"
$a^2$ errors in the unimproved results.  Unfortunately, the large
errors preclude definitive conclusions. Nevertheless, the fact that
our result $0.58(4)$ differs from the ``$\alpha^2$ removed" unimproved
result of $0.76(6)$ by about $2\sigma$, while lying closer to the
continuum result $0.62(4)$, is consistent with our conclusion that
discretization errors are reduced by using smeared links.

\section{Conclusion}
\label{sec:conclude}

We conclude that improved staggered fermions are a viable and
promising option for calculations of $m_s$ and $B_K$ in full QCD
simulations. The difficulties observed with unimproved staggered
fermions (ill behaved perturbation theory for $Z_m$ in the case of
$m_s$ and discretization errors in $B_K$) are greatly reduced. Our
study suggests that reliable calculations should be possible on the
ensembles of lattices being generated with dynamical improved
staggered fermions without requiring very small lattice
spacings. There are two caveats, however. To reduce the uncertainty
due to the two-loop term in the renormalization constants below
our estimates of $4\%$ in $m_s$ and $7\%$ in $B_K$ will require a 
demanding two-loop or
non-perturbative calculation of matching factors and
the calculation of a larger set of lattice matrix elements. Second, in the case
of $B_K$, the estimate in the chiral limit is very sensitive to errors
in the matching factors, as well as to the chiral extrapolation.
%

%
%
%
%\begin{table}[t!]
%\caption{Bare $B_A$ and $B_V$
%\label{tab:BA_and_BV}}
%\begin{ruledtabular}
%\begin{tabular}{c | l l l l}
%$m_q$ & $B_A(I)$ & $B_A(II)$ & $B_V(I)$ & $B_V(II)$ \\
%\hline
%$m_1$ & 2.374(280) & 1.485(64) & -2.735(209) & -0.637(99)  \\
%$m_2$ & 1.193(92)  & 1.131(32) & -1.482(77)  & -0.232(31)  \\
%$m_3$ & 0.758(39)  & 0.981(19) & -0.976(39)  & -0.105(13)  \\
%$m_4$ & 0.559(22)  & 0.907(13) & -0.726(27)  & -0.055(8)  
%\end{tabular}
%\end{ruledtabular}
%\end{table}
%

%-----------------
% Acknowledgement
%-----------------
%Preliminary versions of this work appeared in
%Refs.~\cite{ref:wlee:1,ref:wlee:2}.
% and is part of our staggered $\epsilon'/\epsilon$ project.
%
% 

\section{Acknowledgement}

This calculation has been done on Columbia QCDSP supercomputer.  We
thank N.~Christ, C.~Jung, C.~Kim, G.~Liu, R.~Mawhinney and L.~Wu for
their support on this staggered $\epsilon'/\epsilon$ project.  This
work is supported in part by BK21, by Interdisciplinary Research Grant
of Seoul National University and by KOSEF contract
R01-2003-000-10229-0, the US-DOE grant KA-04-01010-E161 and contract
DE-FG02-96ER40956.
We gratefully acknowledge discussions with A.~Soni. 
%

%-----------
% reference
%-----------

\end{document}